\documentclass[sn-mathphys-num]{sn-jnl}

\usepackage{xcolor}
\usepackage{graphicx}%
\usepackage{multirow}%
\usepackage{amsmath,amssymb,amsfonts}%
\usepackage{amsthm}%
\usepackage{mathrsfs}%
\usepackage[title]{appendix}%
\usepackage{xcolor}%
\usepackage{textcomp}%
\usepackage{manyfoot}%
\usepackage{booktabs}%
\usepackage{algorithm}%
\usepackage{algorithmicx}%
\usepackage{algpseudocode}%
\usepackage{listings}%
\usepackage{dsfont}%
\usepackage{graphicx}%
\usepackage{amsmath}%
\usepackage{float}%
\usepackage{subfloat}%
\usepackage[inline]{enumitem}%

\theoremstyle{thmstyleone}%
\theoremstyle{thmstyletwo}%
\newtheorem{conjecture}{Conjecture}

\theoremstyle{thmstylethree}%

\newcommand{\ket}[1]{ | \, #1 \rangle} \newcommand{\bra}[1]{ \langle #1 \, |}

\newcommand{\bk}[2]{ \langle #1 | #2 \rangle}
\newcommand{\TN}[1]{ \left| \left| #1 \, \right| \right|_1} 
\newcommand{\Norm}[1]{ \left| \left| #1 \, \right| \right|} 
\newcommand{\Ab}[1]{ \left| #1 \, \right|} 

\DeclareMathOperator{\Tr}{Tr}
\DeclareRobustCommand\openone{\leavevmode\hbox{\small1\normalsize\kern-.33em1}}%

\begin{document}

\title[How decoherence affects the security of BB84 quantum key distribution protocol]{How decoherence affects the security of BB84 quantum key distribution protocol}


\author*[1]{\fnm{Robert} \sur{Okuła}}\email{rbrt.okula@gmail.com}

\author[1,2,3]{\fnm{Piotr} \sur{Mironowicz}}\email{piotr.mironowicz@gmail.com}

\affil[1]{\orgdiv{Department of Algorithms and Systems Modelling, Faculty of Electronics, Telecommunications and Informatics}, \orgname{Gdańsk University of Technology}, \orgaddress{\street{Narutowicza 11/12}, \city{Gdańsk}, \postcode{80-233}, \country{Poland}}}

\affil[2]{\orgdiv{Department of Physics}, \orgname{Stockholm University}, \orgaddress{\street{Roslagstullsbacken 21}, \city{Stockholm}, \postcode{114 21}, \country{Sweden}}}

\affil[3]{\orgdiv{International Centre for Theory of Quantum Technologies}, \orgname{University of Gdańsk}, \orgaddress{\street{Jana Bażyńskiego 1A}, \city{Gdańsk}, \postcode{80-309}, \country{Poland}}}


\abstract{
We present how the mechanisms of quantum Darwinism allow for the leakage of information in the standard BB84 quantum key distribution protocol, a paradigmatic prepare and measure quantum cryptography scenario.

We work within the decoherence theory framework and employ the model of measurements provided by quantum Darwinism. We investigate how much of the information about the results crucial for the cryptographic key to be kept secret is leaked during the quantum measurement process and subsequently how much of that information might be later obtained by an eavesdropper using a type of a so-called Van Eck side-channel wiretapping.

We also show how the security can be affected by different ways of organizing the surrounding environment into layers, e.g. rooms or other divisions affecting the spread of quantum information in the environment and its interaction, paving a venue to potential enhancements, and insight into proper engineering of shieldings for cryptographical devices.}




\maketitle

\newpage

\section{Introduction}
\label{sec1}

Quantum mechanics stands as a cornerstone of modern physics, representing a paradigm shift that has profoundly altered our understanding of the universe. Its principles, often defying classical intuition, have reshaped our worldview, challenging conventional notions of reality and opening new frontiers of exploration.

The practical ramifications of quantum mechanics extend far beyond the realm of theoretical physics, permeating into diverse fields of application. Quantum cryptography~\cite{pirandola2020advances}, exemplified by applications like quantum key distribution (QKD)~\cite{renner2008security}, represents one such domain where the principles of quantum mechanics are harnessed to enhance security. These cryptographic protocols offer unparalleled levels of security compared to their classical counterparts, leveraging the inherent properties of quantum systems to safeguard sensitive information.
Among the pioneering protocols in quantum cryptography is the Bennett Brassard protocol from 1984 (BB84)~\cite{bb84}, which laid the foundation for secure communication in the quantum realm. By exploiting the quantum properties of photons, BB84 facilitates the exchange of cryptographic keys with provable security, thereby ensuring the confidentiality of transmitted data.
Central to the implementation of quantum cryptographic protocols is the assumption of a secure laboratory environment, where the classical processing of data from quantum devices remains under controlled conditions. This foundational assumption is essential for preserving the integrity of quantum communication channels and mitigating the risk of unauthorized access to sensitive information unless transmitted by an unsecured channel.

However, the enigmatic nature of quantum mechanics is underscored by the measurement problem, a fundamental quandary concerning the irreconcilable disparity between the unitary, reversible evolution of quantum states and the irreversible nature of measurement~\cite{Schlosshauer_2005}. Wigner's friend paradox encapsulates this conundrum, highlighting the profound philosophical implications of quantum measurement theory~\cite{wigner1963problem,deutsch1985quantum}.
In seeking to address the mysteries surrounding quantum measurement, the theory of quantum Darwinism (QD)~\cite{Zurek_2009}, formulated by Żurek in the 1980s~\cite{PhysRevD.24.1516,PhysRevD.26.1862}, has garnered significant attention. QD posits that the role of the environment surrounding a quantum system is crucial in the process of measurement, with information becoming intersubjective and classical through its widespread dissemination~\cite{Ajdukiewicz1977,mironowicz2018system}.
\textit{Decoherence}, a phenomenon arising from the interaction between a quantum system and its environment, lies at the heart of QD. This process, essential for the emergence of classical behavior, results in the loss of the so-called coherence between degrees of freedom of a physical object, and the establishment of classical-like states, thereby enabling the manifestation of measurement outcomes~\cite{zurek2003decoherence}.

The recognition of the environment's critical role in quantum measurement and decoherence opens the door to novel security threats that have not been thoroughly explored~\cite{Mironowicz_2022}. Analogous to van~Eck attacks in classical computer science and networks, where electromagnetic emanations from electronic devices are intercepted to glean sensitive information, quantum systems may be susceptible to similar vulnerabilities~\cite{van1985electromagnetic}. The so-called van~Eck attacks exploit unintended emissions from electronic devices, such as computer monitors and other equipment, to reconstruct the concealed information remotely using uncontrolled side-channels ~\cite{standaert2010introduction}, which are often easily omitted in the security analysis of protocols~\cite{messerges1999investigations,loughry2002information,kuhn2005security,tanaka2007information,de2007differential,maiti2019light,morales2020digital,nassi2021glowworm,lavaud2021whispering,lee2022quantitative}. Thus, it is important in the context of quantum devices to investigate whether the vulnerabilities stemming from environmental interactions could be exploited to extract quantum information and thus compromise cryptographic protocols, posing security risks in quantum communication.

To this end, we need to recognize the environment's involvement in quantum measurement necessary for the functioning of quantum devices. As stated above, given the indispensable role of measurement in quantum systems, every quantum device relies on environmental interactions to produce observable outcomes. Consequently, the widespread of quantum information to the environment is necessary for measurement and the emergence of classical behavior~\cite{paz2002environment,Zurek_2004,ollivier2004objective}. This renders quantum devices susceptible to external manipulation and unauthorized access, as the environmental interactions occur beyond the confines of controlled laboratory settings. From this, it seems that quantum devices are inherently vulnerable to attacks analogous to van~Eck attacks in classical systems. In consequence, ensuring the security and integrity of quantum communication requires robust measures to safeguard against potential threats arising from environmental interactions and the dissemination of quantum information.

The article is structured as follows: in sec.~\ref{sec2} we introduce crucial concepts that our study is based upon, namely the description of decoherence from the QD, prepare and measure quantum key distribution, as well as Helstrom state discrimination. This is subsequently used in the discussion in sec.~\ref{sec3}, where we present analytical results for the key rate of single-qubit layers and the numerical analysis of distinguishability.

\section{Methods}
\label{sec2}

We describe the concept of QD, focusing on the information dissipating aspect of the decoherence process, in sec.~\ref{ssec:QD}. In sec.~\ref{ssec:QKD} we shortly describe prepare-and-measure quantum key distribution, to tie these two concepts together in sec.~\ref{ssec:MBB84}, where we describe decoherence in the communication in the quantum channel for BB84 protocol. In sec.~\ref{ssec:Helstrom} we describe the Helstrom measurement used in the security analysis conducted further in this work.

\subsection{Quantum Darwinism}
\label{ssec:QD}

As mentioned above, one of the core ideas that encapsulate the specifics of quantum mechanics is the process of the transition from quantum data to classical, known as quantum measurement~\cite{Schlosshauer2007}. It remains, however, in many ways an obscure phenonemon~\cite{doi:10.1126/science.1109541}, with plenty of approaches that try to explain it, one of which is the decoherence theory, which attempts to explain the loss of quantum characteristics by the system in the lack of full isolation, via the loss of information to the environment either induced by the intentional measurement or the naturally occurring interaction with the environment~\cite{Schlosshauer_2005}.

To better understand the process of decoherence, the need for a method of description and quantification of the amount of information outflow becomes apparent, as pioneered by the groundbreaking works of Wojciech Żurek published in the early 1980s~\cite{PhysRevD.24.1516,PhysRevD.26.1862}. In~\cite{PhysRevD.24.1516} Wojciech Żurek elucidated the process occurring between certain parts of an interacting ensemble aiming to explain how the basis of a particular measurement is formed. The quantum \textit{system} that is a point of interest of a particular observation or measurement is in this approach examined via an \textit{apparatus}, which establishes a nonseparable correlation with the initial system. The process in which the observed system interacts with the measuring apparatus, causing the system to become entangled with its degrees of freedom, is called \textit{premeasurement}~\cite{herbut2014review}.

The research indicated that in the natural process, the eigenvectors of the complete set of observables that commute with the Hamiltonian operator that describes the interaction between the apparatus and the environment can be pointed out as the basis in which the resultant (classical) state will be set – it was named a \textit{pointer basis}. The \textit{environment-induced superselection}, or \textit{einselection}, is the process of designating thas specific preferred basis. The basis has a low vulnerability to further decoherence, making it robustly encoded in the environment. As a result, the system performs effectively classically, exhibiting distinct qualities that are stable and easily observable without major disturbance. This is achieved by suppressing the coherence between different states of the pointer basis. It contributes to the understanding of why, although being essentially made up of quantum particles, macroscopic objects seem to have clearly defined classical features~\cite{zurek1998decoherence,zurek2000einselection}. This created the foundations for QD, which additionally elucidates the methodology for the quantification of the information that spreads through the process.

Since coherence is a quantum phenomenon, then whenever a measurement, which is a process of transiting from quantum to classical data, occurs, it is also necessary that the decoherence takes place~\cite{Zurek_2009}.
The \textit{decoherence factor} is one of the quantities expressing the level of decoherence between a set number of subsystems, which are indexed by $k$. It is defined as the norm of the off-diagonal terms between the observed system's pointer states.~\cite{PhysRevD.24.1516} We utilize it in the form of collective decoherence factor~\cite{PhysRevLett.118.150501}
\begin{equation}
	\label{eq:collectiveDecoherenceFactor}
    \Gamma = \sum_{i \ne j} \left(\Ab{\sigma_{ij}} \prod_{k=1}^{(1-f)M} \Ab{\gamma_{ij}^{(k)}} \right),
\end{equation}
where $\sigma_{ij} = \bra{i} \rho_0 \ket{j}$ and $\gamma_{ij}^{(k)} = \Tr \left(\rho_{i, j}^{(k)} \right) = \Tr \left( U_i^{(k)} \rho^{(k)}_0 U_j^{(k)\dagger}\right)$, where $U_i^{(k)}$ is a unitary evolution operator on the subsystem $k$ conditioned upon the observed system; $\ket{i}$, $\ket{j}$ are pointer states, and $\rho_0$ is the initial state. For full decoherence we have $\Gamma \approx 0$, and when $\Gamma \approx 1$ decoherence did not occur at all.

\subsection{Quantum Cryptography and Prepare and Measure Protocols}
\label{ssec:QKD}

Quantum cryptography is one of the most well-proliferated usages of the quantum information theory. From that, probably the most successful part is QKD, which allows for communication using symmetric encryption with the key securely established between parties with the help of the quantum paradigm. These protocols can be further classified into essentially two types: entanglement-based~\cite{ekert1991quantum,barrett2005no} and prepare-and-measure based~\cite{bb84,bennett1992experimental,bennett1992quantum,pawlowski2011semi}. In this work, we concentrate on the latter type.

The BB84 protocol~\cite{bb84}, presented in 1984, still remains a primary method of its kind. The first party, Alice, generates, using an unbiased random number generator, a string of bits. The string corresponds to the bits that should be transmitted to the second party, Bob, in the key exchange process. Additionally, for every bit transmitted, Alice should randomly choose a defined, orthonormal base randomly and depending on that random choice, prepare a qubit that corresponds to the value of the random transmitted bit ($\ket{0}$ and $\ket{+}$ often correspond to key bit zero and $\ket{1}$ and $\ket{-}$ to key bit one). This qubit is subsequently transferred to Bob using an authenticated quantum channel. The transmission can be potentially manipulated by an eavesdropper or disturbed by the noise. Later on, Bob randomly picks a measurement basis (computational or Hadamard’s) and performs a measurement. The observable corresponds to one of the binary values. However, the basis he used for the measurement can be different than Alice's. In this case, the probability distribution (because they use mutually unbiased bases~\cite{durt2010mutually}) of receiving any of the values will be uniform and no meaningful information will be obtained. Both parties need to check which bases have been used, so they announce that in the unprotected classical channel. If for the $n$-th bit, the measurement bases are different, the parties simply reject that bit.

\subsection{Measurement and Decoherence in BB84 Protocol}
\label{ssec:MBB84}

We focus on the measurement process as established in the BB84 protocol. We analyze the physical exchange as if it were done qubit-by-qubit, without considering potential memory effects of the devices~\cite{masanes2014full}. That allows us to break it down into the pieces that are crucial for studying the behavior of the measurement process. In essence, it is a one-by-one qubit transmission, even if it is later effectively bundled into a full key. Some of these exchanges are going to be rejected, viz. those in which Alice and Bob used different bases for bit encoding and qubit measuring. For every set of bases we have then two bits that need to be distinguished by the eavesdropper. Based on that in BB84 we can show $4$ cases of qubit exchange scenarios that can be analyzed and deconstructed by our model.
Following~\cite{PhysRevD.24.1516} we call the observed qubit a system ($S$), the measuring device an apparatus ($A$), and the third subsystem the environment ($E$).
In~\cite{PhysRevLett.118.150501}, we established that measurement consists of decoherence and orthogonalization. Here, we focus solely on decoherence, assuming full distinguishability of the states received by Bob, as noise in the transmission channel is beyond the scope of our investigation in this work.

Modeling of interaction between every subsystem depends on the choice of the measuring basis. First, we model the interaction between the system and the apparatus similarly to~\cite{Mironowicz_2022}, which is based on specific rank-1 projectors and a CNOT operation. However, we are also taking into account the fact that the premeasurement can be done using two mutually unbiased observables with projectors $\{P_0^{(S)}, P_1^{(S)}\}$ or $\{P_+^{(S)}, P_-^{(S)}\}$ and a CNOT operation is also defined in connection with the measurement basis. The process of premeasurement is therefore described by these unitary operators:
\begin{equation}
    \label{eq:USA_comp}
    U^{(SA)}_{comp} = P_0^{(S)} \otimes \mathds{1}_2^{(A)} + P_1^{(S)} \otimes C_X^{(A)}(\alpha),
\end{equation}
\begin{equation}
    \label{eq:USA_had}
    U^{(SA)}_{had} = P_+^{(S)} \otimes \mathds{1}_2^{(A)} + P_-^{(S)} \otimes C_Z^{(A)}(\alpha),
\end{equation}
where $C_X$ and $C_Z$ operations are defined as:
\begin{align}
	C_X = 
	\begin{bmatrix}
		\sin(\alpha) & \cos(\alpha)\\
		\cos(\alpha) & -\sin(\alpha)
	\end{bmatrix}, \\
	C_Z = H C_X H,
\end{align}
with $\alpha$ set as a constant value. In other words, this transformation establishes a correlation between the observed state and the apparatus, which is crucial, as the reality is observed based on the output values on the apparatus devices, but not directly. This correlation is therefore what determines the course of the measurement – if no information could be transferred between $S$ and $A$, Bob would not be able to obtain any results. This process is defined in relation to the type of measurement that Bob conducts: if he (randomly) chooses the computational basis,~\eqref{eq:USA_comp} applies, and if he measures in Hadamard's basis otherwise,~\eqref{eq:USA_had} is going to apply.

To describe the interaction between the apparatus and the environment, we first need to establish the structure of the environment itself. In this work we assume that the environment consists of $N_l$ layers of $N_E$ qubits each, with the initial state of the environment (tensor multiplied by the initial states of the system and apparatus):
\begin{align}
	\label{eq:initial_env}
	\ket{\phi}_{0, comp}^{(SAE)} = \ket{\psi}^{(S)} \otimes \ket{0}^{(A)} \otimes \bigotimes_{l = 1}^{N_l} \bigotimes_{e = 1}^{N_E} \ket{0}^{(E_{l,e})}, \\
	\ket{\phi}_{0, had}^{(SAE)} = \ket{\psi}^{(S)} \otimes \ket{+}^{(A)} \otimes \bigotimes_{l = 1}^{N_l} \bigotimes_{e = 1}^{N_E} \ket{+}^{(E_{l,e})}.
\end{align}
The interaction between the apparatus and the environment is for the purspose of our considerations defined as:
\begin{equation}
	\label{eq:UAE}
	\begin{split}
		U^{(AE)} = \left(P_0^{(E_{N_l - 1})} \otimes U_0^{(E_{N_l})} + P_1^{(E_{N_l - 1})} \otimes U_1^{(E_{N_l})}\right) \times \\
		\vdots \\
  		\left(P_0^{(E_1)} \otimes U_0^{(E_2)} + P_1^{(E_1)} \otimes U_1^{(E_2)}\right) \times \\
        \left(P_0^{(A)} \otimes U_0^{(E_1)} + P_1^{(A)} \otimes U_1^{(E_1)}\right).
	\end{split}
\end{equation}
For the sake of simplicity, we assume that the interactions occur subsequently, that is the interaction with the apparatus and the first layer (the last multiplication factor) of the environment occurs, then between the first and the second layer, and the like. The first multiplication factor from last in $U^{(AE)}$ describes the interaction of the state $A$ of the apparatus, that in result modifies the state of the subsystem of the first environmental layer conditioned on the decoherence-based projectors on the antecedent subsystem. Subsequent multiplication factors are constructed likewise for successive pairs of layers ($E_1$ transfering information to $E_2$, $E_2$ to $E_3$ etc.). This interaction is the last one during the process and leads to decoherence (after the interaction of that kind and tracing out some subsystems the off-diagonal factors are sufficiently low) and information propagation to the last layer of the environment $E_{N_l}$, which is accessible to the eavesdropper. Projectors $P_0^{(X)}$ and $P_1^{(X)}$ ($X \in \{A, E_1, \ldots, E_{N_l}\}$) on the state of the apparatus are regular, single-qubit projectors for Hadamard and computational bases. In the numerical simulation, the interaction between the layers is based on the Haar distributed $U_0^{E_l}$ and $U_1^{E_l}$. Projectors on the preceding layers are constructed as Hermitian matrices that are parameterized using randomly chosen real matrices, to take into account the unpredictability of the free-space quantum objects interacting with each other.
The model we propose here is a simplified toy-model, what, due to high complexity of of any quantum description of classical objects, is a standard approach, see e.g.~\cite{cucchietti2005decoherence,touil2022eavesdropping}.

As the bases used for encoding and measurement are publicly known, the eavesdropper has to only distinguish between two cases – if the key bit was $0$ or $1$. These two bits in the process will result in different states of the environment in the end and there is only one (last or the only one) layer of the environment that Eve has access to. Therefore, she needs to prepare a measurement that can distinguish between two states made of $N_E$ qubits.

The spread and accessibility of the information can be however constrained by noise and shielding which needs to be modeled properly.
Additionally, only part of that layer might be distinguishable, thus the projector that the eavesdropper uses on her available layer should be an arbitrary projector of a rank $2^k$, where $k$ denotes the number of qubits that the eavesdropper can distinguish using the Van Eck-type antenna~\cite{VANECK1985269}. In this type of attack, so-called Van Eck phreaking, the eavesdropper can pick up side-channel information. In our scenario, this side information is contained in the environmental qubits.

\subsection{Helstrom state discrimination}
\label{ssec:Helstrom}
The eavesdropper's main goal is to distinguish between two different resultant states procured in the process mentioned in sec.~\ref{ssec:MBB84}. This can be achieved with Helstrom measurement~\cite{Helstrom1969}, which is the measurement that leads to the lowest possible error in distinguishing between two states. The probability of a correct guess is given by the Holevo–Helstrom theorem~\cite{HOLEVO197333}:
\begin{equation}
	P_{guess} = \frac{1}{2} + \frac{1}{2} \TN{ \lambda \rho_0 - (1 - \lambda) \rho_1 },\label{eq:hht}
\end{equation}
where $\lambda$ is the probability of appearance of the first state (for uniform probability distribution $\lambda = 0.5$) and $\TN{\cdot}$ is a trace norm.

\section{Results}
\label{sec3}

In sec.~\ref{sec3.1} we discuss an analytical model fo the environmental information spread demonstrating a formula indicating the key rate in a particular class of scenarios, whereas in sec.~\ref{sec3.2} we present the results of a numerical simulation.

\subsection{Single-qubit layers}
\label{sec3.1}

We have created a simplified model of the above situation, in which we have $N_l$ of single-qubit layers ($N_E = 1$). As for BB84, we only consider the situations in which the key bit is accepted by both parties, as this can be observed in the classical, unencrypted, channel. The base that was used in the communication process is also shared, therefore the eavesdropper only has to distinguish if the state of the last (available) layer of the environment (subsystem $E_{N_l}$) indicates the leak of the information about the quantum state suggesting that the exchanged key bit was 0 or suggesting that it was 1. Therefore, Eve has to make a Helstrom measurement~\cite{Helstrom1969}. The probability of success of this type of measurement is given by~\eqref{eq:hht}. We make a common assumption that Alice is often said to be using a near-perfect random bit generator, so we can assume that $\lambda = 0.5$.

To model the interactions constituting $U^{AE}$ as in~\eqref{eq:UAE}, that involve the interaction between the apparatus and the first layer, as well as between the layers, we need first to include the noise effect between the layers into an imperfect CNOT operator, with the imperfection defined as the rotation of a regular CNOT operator by angle $\alpha$. To achieve that, we introduce the operator $Q(\varepsilon, \alpha) = \varepsilon \cdot \mathds{1}_2 + (1-\varepsilon) \cdot C_X(\alpha)$ (similarly for $C_Z$), where coefficient $\varepsilon$ parametrize the interaction between the apparatus and the environment (and also between subsequent layers) that limits the degree of interaction between two subsystems, see~\eqref{eq:noisedInteraction} below. However, $Q$ is not unitary, therefore it must be transformed into a unitary operator. One way to perform such transformation will be to use some kind of matrix factorization that as a result provides a unitary matrix (preferably as close to the original as possible) and some additional product, which we can reject. QR factorization meets the requirement, as it decomposes any real square matrix into a product $Q = Q'R$, where $Q'$ is an orthonormal matrix and $R$ is an upper triangular matrix, and $R = \openone$ when $Q$ is unitary. Using this method (the whole procedure is described in Appendix 1) we obtain:
\begin{equation}
    \label{eq:noisedInteraction}
	Q'(\varepsilon, \alpha) =
	\begin{bmatrix}
		\frac{p}{\sqrt{p^2 + q^2}} & \frac{-q}{\sqrt{p^2 + q^2}}\\
		\frac{q}{\sqrt{p^2 + q^2}} & \frac{p}{\sqrt{p^2 + q^2}}
	\end{bmatrix},
\end{equation}
where $p = \varepsilon + (1-\varepsilon) \sin(\alpha)$ and $q = (1-\varepsilon) \cos(\alpha)$. With $Q'$ we now define the interlayer unitary interaction operator:
\begin{equation}
	\label{eq:interlayerCNOT}
    U = P_0^{(X)} \otimes \mathds{1}_2^{(Y)} + P_1^{(X)} \otimes Q'(\varepsilon, \alpha)^{(Y)},
\end{equation}
where $X$ and $Y$ are consecutive environmental layer subsystems. Subsequently, we get from direct calculations that the apparatus value of the guessing probability for the single-layer environment is given by
\begin{equation}
	\label{eq:Pguess1}
    P_{guess}^{N_l = 1} = \frac{1}{2} + \frac{1}{2} \cdot \frac{q}{\sqrt{p^2 + q^2}} 
\end{equation}
and for the two-layer environment
\begin{equation}
	\label{eq:Pguess2}
    P_{guess}^{N_l = 2} = \frac{1}{2} + \frac{1}{2} \cdot \frac{q^3}{(p^2 + q^2)^2\sqrt{p^2 + q^2}}.
\end{equation}
The eavesdropper is conducting the measurement on the last layer $E_{N_l}$, as this is the only layer he have access to, and we assume the worst-case scenario that he can obtain the information that dissipated to that layer fully. Thus, the mutual information between Bob's apparatus and the last layer $I(A:E_{N_l})$ should be directly related to $P_{guess}$. For instance, when the eavesdropper receives the bit 0, he can be sure with the probability $P_{guess}$ that it is the right one, and with the probability $1 - P_{guess}$ that it is the wrong one. In other words, $p(e = 0 | a = 0) = P_{guess}$ and $p(e = 0 | a = 1) = 1 - P_{guess}$ and similarly for bit 1. This gives us the conditional entropy $H(E_{N_l} | A)$, which is equal to $H(A | E_{N_l})$, as the situations in which $a = 0$ conditioned on $e = 0$ can also only happen when Eve guesses correctly, therefore $p(a = 0 | e = 0) = P_{guess}$. Additionally, $H(A) = 1$ in this case, because Alice is using a random number generator with an approximately uniform probability distribution.

This leads us to the general form for mutual information:
\begin{equation}
	\begin{aligned}
	    I(E_{N_l}:A) &= H(A) - H(A|E_{N_l})\\
	    &= 1 + (1-P_{guess}) \log_2 (1-P_{guess}) + P_{guess} \log_2 (P_{guess}).
	\end{aligned}
\end{equation}
Mutual information is therefore dependent solely on the value of $P_{guess}$. I. Csiszar and J. Körner~\cite{1055892} showed that parties can obtain a secret key only if the mutual information between the parties is larger than between Alice and Eve and the key rate is the difference between these two values: $r = I(S:A) - I(S:E_{N_l})$. We are considering only the situations in which communication leads to the accepted bit of the key, therefore $I(S:A) = 1$ and $I(S:E_{N_l}) = I(E_{N_l}:A)$, thus the general formula for key rate is dependent only on the value of $P_{guess}$ as well:
\begin{equation}
	\label{eq:keyrate}
    r = (1-P_{guess}) \log_2 (1-P_{guess}) + P_{guess} \log_2 (P_{guess}).
\end{equation}
For example, the key rate for the single-qubit layer with exactly one layer environment is given as:
\begin{equation}
    r_{N_l = 1} = 1 - \frac{1}{2} \left(\log_2 \frac{q^2}{p^2 + q^2} + \frac{q}{\sqrt{p^2 + q^2}} \log_2 \frac{\sqrt{p^2 + q^2} + q}{\sqrt{p^2 + q^2} - q}\right).
\end{equation}
We have calculated several additional examples for different numbers of layers $N_l$, and based on the observation for their form, cf.~\eqref{eq:Pguess1} and~\eqref{eq:Pguess2}, as well as the general form of key rate~\eqref{eq:keyrate} we conjecture the following for single-qubit layers:
\begin{conjecture}\label{con1}
	 The guessing probability value, $P_{guess}$, for $n$ single-qubit layers is
	\begin{equation}
	    P_{guess}^{N_l = n} = \frac{1}{2} + \frac{1}{2} \cdot \frac{q^{2n-1}}{(p^2 + q^2)^{(2n-1)/2}},
	\end{equation}
	and the key rate for $n$ single-qubit layers is
	\begin{equation}
		\begin{aligned}
		    r_{N_l = n} = 1 &- \frac{1}{2} \cdot \log_2 \frac{(p^2 + q^2)^{2n-1} - q^{2(2n-1)}}{(p^2 + q^2)^{2n-1}}\\
		    &+ \frac{1}{2} \cdot \frac{q^{2n-1}}{(p^2 + q^2)^{(2n-1)/2}} \log_2 \frac{(p^2 + q^2)^{(2n-1)/2} + q^{2n-1}}{(p^2 + q^2)^{(2n-1)/2} - q^{2n-1}}.
		\end{aligned}
	\end{equation}
\end{conjecture}

\subsection{Numerical analysis of guessing probability for multi-qubit layers}
\label{sec3.2}

To investigate how the distinguishability decreases (and thus key rate increases) on different levels of $\varepsilon$ noise between the layers we have created a model and a numerical simulation of the course of the protocol, maintaining appropriate quantum measurements, in the standard BB84 procedure, in which the accepted cases are the ones that keep the same encoding and measurement bases, either Hadamard's or computational. In the simulation, we calculated appropriate operators $U^{(SA)}$ and $U^{(AE)}$ based on their general forms~\eqref{eq:USA_comp} and~\eqref{eq:UAE} and applied them to initial states in a form of~\eqref{eq:initial_env}. Subsequently, we trace out all unneeded subsystems (all but $E_{N_l-1}$). Such result is prepared for both BB84 situations – when the initial, transported, state is $\ket{0}$ (or $\ket{+}$) and when it is $\ket{1}$ (or $\ket{-}$). Having two different results of this preparation, we can calculate $P_{guess}$ based on~\eqref{eq:hht}.

\begin{figure}[htbp]
	\includegraphics[width=\linewidth]{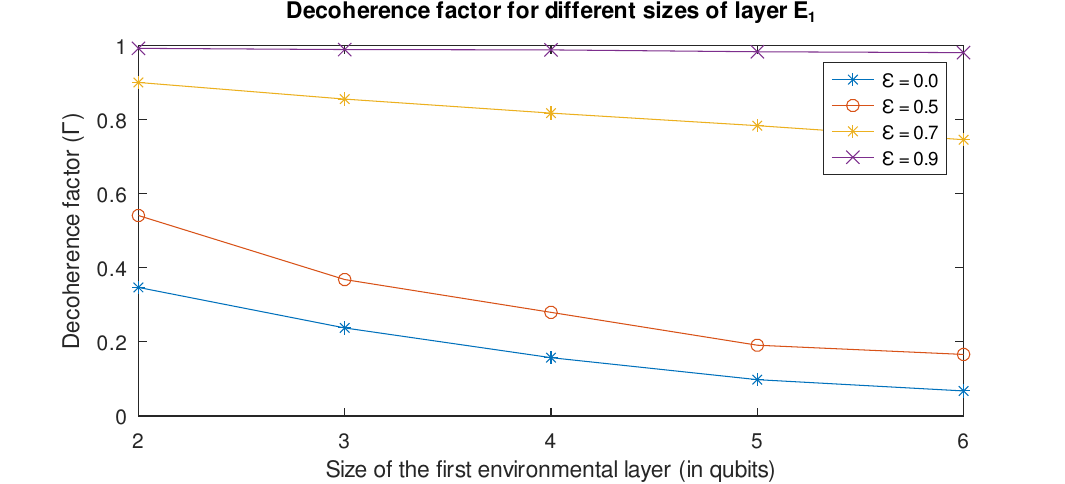}
	\caption{(color online) The dependence of the average decoherence factor $\Gamma$ for the interaction $U^{(AE_1)}$ given the size of the first environmental layer $E_1$. The average is calculated over the cases rejected in BB84, for different interaction degree values ($\varepsilon$). Only the first decoherence interaction, between the apparatus and the first layer of the environment is taken into consideration, as this is where the collapse of the state of the apparatus occurs.}
	\label{fig:decoherence}
\end{figure}

To show the decoherence occurring in the model, we executed numerical simulations for the rejected cases of BB84, where the encoding and measurement are done using different, mutually unbiased, bases. The results of the process in terms of the average collective decoherence factor~\eqref{eq:collectiveDecoherenceFactor}~\cite{PhysRevLett.118.150501} (on both rejected cases) are presented in Fig.~\ref{fig:decoherence}.

We executed the simulation for several qubits in the single environmental layer $N_E$ between $3$ and $7$. For each case, we prepared optimal projectors as Helstrom measurements and calculated guessing probability using the Holevo–Helstrom theorem~\cite{Helstrom1969,HOLEVO197333}. We obtained that the guessing probability changes given the percentage of controlled (and measured) dimensions of the (last or only) layer of the environment (that increases two-fold with every additional qubit). The results are presented in Tab.~\ref{tab:comphad}. Additionally, it can be noticed that the guessing probability for the corresponding degree of control of the last layer increases for the higher number of qubits in the layer – e.g. the $50\%$ dimensionality control of the $N_E = 3$ qubit layer gives us the probability of around $72\%$, but for $N_E = 7$ it is $77\%$. The observation that a larger environment is capable of accommodating more information and compensating for noise has been noted, for example, in~\cite{zwolak2009quantum,zwolak2010redundant}.

\begin{table}[htbp]
	\begin{tabular}{|l|l|l|l|l|l|}
		\hline
		\% of $E$ controlled & $N_E = 3$ & $N_E = 4$ & $N_E = 5$ & $N_E = 6$ & $N_E = 7$ \\ \hline
		1.5625              & -         & -         & -         & -         & 0.509     \\ \hline
		3.125               & -         & -         & -         & 0.513     & 0.521     \\ \hline
		6.25                & -         & -         & 0.518     & 0.527     & 0.533     \\ \hline
		12.5                & -         & 0.556     & 0.556     & 0.563     & 0.565     \\ \hline
		25                  & 0.589     & 0.602     & 0.599     & 0.625     & 0.637     \\ \hline
		50                  & 0.717     & 0.723     & 0.731     & 0.723     & 0.773     \\ \hline
		100                 & 0.967     & 0.983     & 0.995     & 0.997     & 0.998     \\ \hline
	\end{tabular}
	\caption{Guessing probability of a key bit for eavesdropper that uses Helstrom measurement on a controlled part of the environment's sole ($N_l = 1$) layer for $\varepsilon = 0$.}
	\label{tab:comphad}
\end{table}

We observe that the guessing probability is linearly proportional to the controlled part of the environment, counted as the percentage of dimension, and thus the control of the $N_E - 1$ qubits leads to the guessing probability of around 75\%.

The decoherence factor for cases where the key bit is successfully established equals $\Gamma = 0$, as with the encoding and measurement done in the single, same bases, the state $S$ does not decohere, as it is from the beginning a pointer state and not a mixture of pointer states. However, the information transfer still occurs between the apparatus and the environment, as the process of subsequent interactions~\eqref{eq:UAE} between layers still occurs.

\begin{figure}[htbp]
	\centering
	\includegraphics[width=0.9\linewidth]{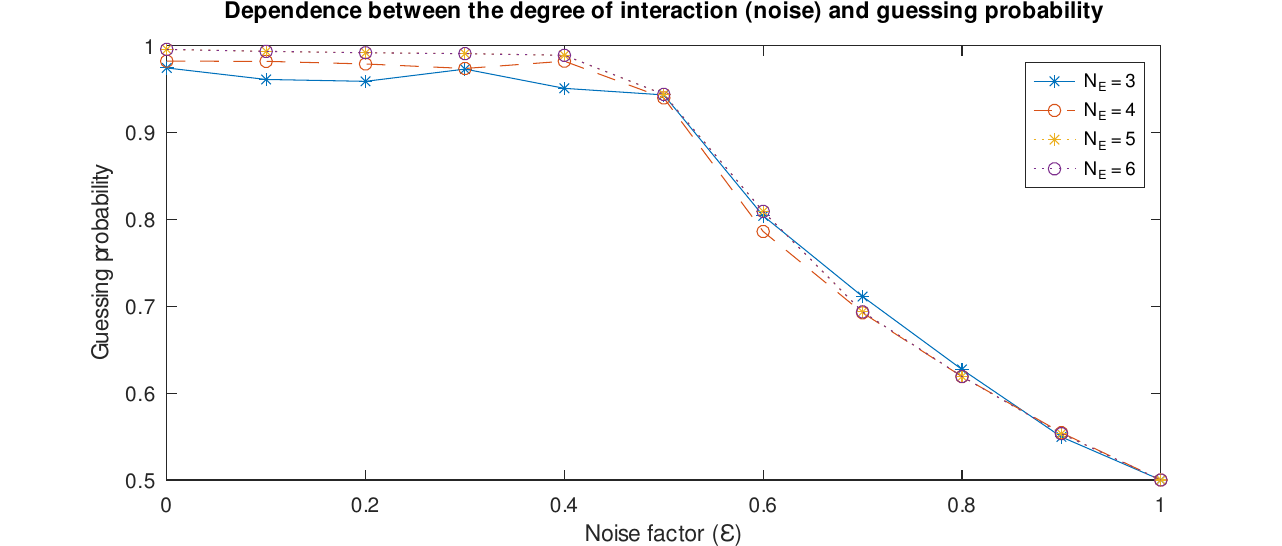}
	\caption{(color online) Guessing probability of a key bit for different values of $\varepsilon$ degree of interaction~\eqref{eq:interlayerCNOT} between apparatus and the environment for a single-layer environment~\eqref{eq:UAE}. The eavesdropper fully controls the last (and only) layer of the environment fully.}
	\label{fig:5qubits}
\end{figure}

The next results are presented in Fig.~\ref{fig:5qubits} and Tab.~\ref{tab:4epsilons}. First, from Fig.~\ref{fig:5qubits} it can be observed that the visible loss of distinguishability i.e. the guessing probability for the eavesdropper in the considered sizes of environments occurrs for $\varepsilon \geq 0.5$, with the monotonic decrease. For $\varepsilon = 1$, there is understandably no interaction between the apparatus and the environment and thus no distinguishability yet also no decoherence, as we noted from Fig.~\ref{fig:decoherence} above. When additional layers are present, a visible decrease of distinguishability occurs between the first and the second layer, with a much smaller loss between the second and the third one as shown in Tab.~\ref{tab:4epsilons}. That difference is also lower for higher values of $\varepsilon$, as $P_{guess}$ overall decreases for higher values of $\varepsilon$.

The data presented in Tab.~\ref{tab:comphad} and in Fig.~\ref{fig:5qubits} strongly suggests that the higher number of qubits in environmental bases (including the last one, accessible to the eavesdropper) leads to a higher level of information extraction, even if the controlled percentage of the environment is the same. The guessing probability also fluctuates less for lower values of noise factor $\varepsilon$, as seen in Fig.~\ref{fig:5qubits} for $\varepsilon < 0.4$. This phenomenon is in accordance with the observation of the so-called \textit{information plateau}. The information plateau denotes a precise point in time, independent of the size of the environmental portion under consideration, at which the mutual information between the system and various environmental components achieves a constant value. This constant number indicates that there is classical objectivity because the system's information is redundantly encoded across the environment.~\cite{blume2005simple,Zurek_2009,chen2019emergence,milazzo2019role,ccakmak2021quantum}.

\begin{table}[htbp]
	\begin{tabular}{|l|l|l|l|}
		\hline
		$N_l$ (number of layers) & $\varepsilon = 0.5$ & $\varepsilon = 0.7$ & $\varepsilon = 0.9$ \\ \hline
		1                & 0.871               & 0.658               & 0.549               \\ \hline
		2                & 0.823               & 0.646               & 0.536               \\ \hline
		3                & 0.818               & 0.645               & 0.536               \\ \hline
	\end{tabular}
	\caption{Key bit guessing probability for $N_E = 2$ and different values of $\varepsilon$ degree of interaction between apparatus and the environment and different number of layers ($N_l = 1, 2, 3$).}
	\label{tab:4epsilons}
\end{table}

The results of our numerical calculations offer insights into the dynamics of decoherence and its implications for quantum communication protocols. In particular, our findings suggest that a layered environment serves as a robust model for simulating decoherence phenomena. By examining the behavior of the decoherence factor $\Gamma$, as depicted in Fig.~\ref{fig:decoherence}, we infer that in real-world scenarios where $\Gamma$ is expected to be on the order of $10^{-40}$~\cite{Zurek86}, the presence of a sufficiently large layer would likely be accessible to potential eavesdroppers. 

Moreover, our analysis, as summarized in Tab.~\ref{tab:comphad}, underscores a significant trend: as the size of the layer increases, the less part of the layer needs to be controlled by the eavesdropper to increase guessing probability, consequently elevating the probability of a successful guessing attack on the BB84 protocol. This phenomenon suggests that even a minute fraction of the layer under the control of an eavesdropper could render the protocol vulnerable to security breaches. 

Furthermore, our investigation reveals a nuanced relationship between the size of the layer and the noise parameter $\varepsilon$. As illustrated in Fig.~\ref{fig:decoherence}, we observe that larger layers are necessitated to counterbalance higher levels of noise while keeping the same value of $\Gamma$, as evidenced by the increasing trend of $\varepsilon$. 
Conversely, as shown in Fig.~\ref{fig:5qubits} and in Tab.~\ref{tab:comphad} while a larger $\varepsilon$ corresponds to a reduction in guessing probability, 
the presence of a larger layer confers enhanced robustness against noise, highlighting an intricate trade-off involved in optimizing layer size and noise parameters to mitigate eavesdropping risks effectively. 

Additionally, our results shed light on the potential efficacy of employing multiple layers as a security measure. As indicated in Tab.~\ref{tab:4epsilons}, the utilization of multiple layers may introduce additional difficulty for eavesdroppers, thereby enhancing the difficulty of unauthorized access. This observation underscores the importance of devising sophisticated shielding mechanisms that facilitate necessary measurements while upholding stringent security standards. 

\section{Conclusions}

In our study, we analyzed the levels of eavesdropping enabled by the process of decoherence that can influence exchange of the key between interested parties. We showed how the process of information propagation that is, according to the theory of quantum Darwinism, required during the decoherence of the measured state, increases the eavesdropper's likeliness of correctly guessing the accepted key bit. To model more advanced scenarios of that phenomena, we introduced the concept of environmental layers, which during the joint interaction between the subsequent layers introduce some noise, determined by the constant noise factor $\varepsilon$. Based on that concept, we were able to describe the interaction process between these layers, including a formula for mutual information and the final key rate that is dependent only on the value of guessing probability. We postulated the formula for the interaction between any number of single-qubit layers.

The separate numerical analysis of changes in $P_{guess}$ depending on the parameters of the environment led to the presentation of how the distinguishability (of states of the last layer of the environment that interacted based on the quantum interaction for Alice's bits 0 and 1 respectively) changes depending on the level of noise present in the information dissipation process: the distinguishability falls noticeably with $\varepsilon > 0.5$.
Our numerical analysis reveals that a layered environment effectively models decoherence dynamics, with larger layers posing increased vulnerability to eavesdropping in quantum communication protocols such as BB84. We observe a delicate balance between layer size and noise parameters, with larger layers necessary to counteract higher noise levels. Interestingly, while larger layers enhance the noise resistance (for larger layers and constant $\varepsilon$, $P_{guess}$ is visibly lower), they also exacerbate vulnerability to eavesdropping. Furthermore, employing multiple layers may complicate eavesdropping attempts, highlighting the need for robust shielding mechanisms to ensure both measurement feasibility and security.

The results show the significance of proper shielding against information leakage via the most basic quantum phenomena. An eavesdropper could be able to extract a significant amount of information enough to compromise our communication basically out of thin air. The subsequent research should also focus on providing the methods of security verification, even in the form of new, semi-device independent protocols~\cite{pawlowski2011semi} that mitigate these risks.

\section*{Acknowledgements}
Work partially supported by National Science Centre, Poland, grant number 2018/31/B/ST6/00820, by the Foundation for Polish Science (IRAP project, ICTQT, contract No. 2018/MAB/5, co-financed by EU within Smart Growth Operational Programme), by the Knut and Alice Wallenberg Foundation through the Wallenberg Centre for Quantum Technology (WACQT), and by The National Centre for Research and Development (NCBiR) QUANTERA/2/2020 (www.quantera.eu) an ERA-Net cofund in Quantum Technologies under the project eDICT. The numerical calculations we conducted using GNU Octave and the package QETLAB 0.9.

\begin{appendices}

\section{QR decomposition}
\label{appendixQR}

The incorporation of the noise parameter between the subspaces in the form of $Q(\varepsilon, \alpha) = \varepsilon \cdot \mathds{1}_2 + (1-\varepsilon) \cdot C_X(\alpha)$ leads to a non-unitary matrix, that cannot be used as a quantum operator, therefore we need to decompose it into an orthonormal matrix, with the factor that was causing non-unitarity removed. One approach would be to use the singular value decomposition $M = U \Sigma V^{\dagger}$, where $\Sigma$ represents the rectangular diagonal matrix with singular values on the diagonal, $V$ and $U$ form two sets of orthonormal bases. This decomposition for this matrix leads to eigendecomposition that removes the noise factor $\varepsilon$, which makes the transformation useless.

The QR decomposition $Q = Q'R$ transforms the initial matrix into an orthonormal $Q$ and upper triangular matrix $R$. It can be calculated using the Gram–Schmidt process. We divide our matrix into columns $Q = [q_1 | q_2]$:
\begin{equation}
    q_1 = \begin{bmatrix}
        \varepsilon + (1-\varepsilon) \sin (\alpha)\\
        (1-\varepsilon) \cos(\alpha)
    \end{bmatrix}, q_2 = \begin{bmatrix}
        (1-\varepsilon) \cos(\alpha)\\
        \varepsilon - (1-\varepsilon) \sin (\alpha)
    \end{bmatrix}.
\end{equation}
The norm from the vector $\Norm{a_1} = \sqrt{(\varepsilon + (1-\varepsilon) \sin (\alpha))^2 + (1-\varepsilon)^2\cos(\alpha)^2}$. The first column of the matrix equals
\begin{equation}
    u_1 = \frac{q_1}{\Norm{q_1}} = \begin{bmatrix}
        \frac{\varepsilon + (1-\varepsilon) \sin (\alpha)}{\sqrt{(\varepsilon + (1-\varepsilon) \sin (\alpha))^2 + (1-\varepsilon)^2\cos(\alpha)^2}}\\
        \frac{(1-\varepsilon) \cos(\alpha)}{\sqrt{(\varepsilon + (1-\varepsilon) \sin (\alpha))^2 + (1-\varepsilon)^2\cos(\alpha)^2}}
    \end{bmatrix}.
\end{equation}
The second vector is calculated similarly. First, we need need a projection of $q_2$ on $u_1$: $\text{proj}_{u_1} q_2 = \frac{\bk{u_1}{q_2}}{ \bk{u_1}{u_1}} u_1 = \begin{bmatrix}
         2 (1-\varepsilon) \cos(\alpha)\\
        -2 (1-\varepsilon) \sin (\alpha)
    \end{bmatrix}$. This value is used to calculate the intermediate vector $q'_2 = q_2 - \text{proj}_{u_1} q_2 = \begin{bmatrix}
         -(1-\varepsilon) \cos(\alpha)\\
        \varepsilon + (1-\varepsilon) \sin (\alpha)
    \end{bmatrix}$. Analogously:
\begin{equation}
    u_2 = \frac{q_2}{\Norm{q_2}} = \begin{bmatrix}
        \frac{-(1-\varepsilon) \cos(\alpha)}{\sqrt{(\varepsilon + (1-\varepsilon) \sin (\alpha))^2 + (1-\varepsilon)^2\cos(\alpha)^2}}\\
        \frac{\varepsilon + (1-\varepsilon) \sin (\alpha)}{\sqrt{(\varepsilon + (1-\varepsilon) \sin (\alpha))^2 + (1-\varepsilon)^2\cos(\alpha)^2}}
    \end{bmatrix}.
\end{equation}
This leads to the final forms of $Q'$: \begin{equation}
	Q'(\varepsilon, \alpha) =
 	\begin{bmatrix}
		u_1 & u_2
	\end{bmatrix} = 
	\begin{bmatrix}
		\frac{\varepsilon + (1-\varepsilon) \sin(\alpha)}{\sqrt{(\varepsilon + (1-\varepsilon) \sin(\alpha))^2 + (1-\varepsilon)^2 \cos(\alpha)^2}} & \frac{-(1-\varepsilon) \cos(\alpha)}{\sqrt{(\varepsilon + (1-\varepsilon) \sin(\alpha))^2 + (1-\varepsilon)^2 \cos(\alpha)^2}}\\
		\frac{(1-\varepsilon) \cos(\alpha)}{\sqrt{(\varepsilon + (1-\varepsilon) \sin(\alpha))^2 + (1-\varepsilon)^2 \cos(\alpha)^2}} & \frac{\varepsilon + (1-\varepsilon) \sin(\alpha)}{\sqrt{(\varepsilon + (1-\varepsilon) \sin(\alpha))^2 + (1-\varepsilon)^2 \cos(\alpha)^2}}
	\end{bmatrix}.
\end{equation}

\end{appendices}

\bibliography{sn-bibliography}

\end{document}